# Digital Advertising Traffic Operation: Flow Management Analysis

Massimiliano Dal Mas
*me @ maxdalmas.com*

## Abstract

In a Web Advertising Traffic Operation the Trafficking Routing Problem (TRP) consists in scheduling the management of Web Advertising (Adv) campaign between Trafficking campaigns in the most efficient way to oversee and manage relationship with partners and internal teams, managing expectations through integration and post-launch in order to ensure success for every stakeholders involved.
For our own interest we did that independent research projects also through specific innovative tasks validate towards average working time declared on "specification required" by the main worldwide industry leading Advertising Agency.
We present a Mixed Integer Linear Programming (MILP) formulation for end-to-end management of campaign workflow along a predetermined path and generalize it to include alternative path to oversee and manage detail-oriented relationship with partners and internal teams to achieve the goals above mentioned.
To meet clients' KPIs, we consider an objective function that includes the punctuality indicators (the average waiting time and completion times) but also the main punctuality indicators (the average delay and the on time performance).
Then we investigate their analytical relationships in the advertising domain through experiments based on real data from a Traffic Office. We show that the classic punctuality indicators are in contradiction with the task of reducing waiting times. We propose new indicators used for a synthesize analysis on projects or process changes for the wider team that are more sustainable, but also more relevant for stakeholders. We also show that the flow of a campaign (adv-ways) is the main bottleneck of a Traffic Office and that alternate paths cannot improve the performance indicators.

## Keywords

Digital Advertising, Online Advertising, Trafficking, eCommerce, Customers, Clients, Advertising agencies, Stakeholders, Marketing, Products, Shopping, Display, DEM, Programmatic Adv, Native Adv, Social Adv, Affiliation Adv, Branded Content, Content Recommendation, AdServer, Adv Operations Management, Account Manager, Media planning, Ad Platforms, Industry leading third party Ad Servers, Mixed Integer Linear Programming (MILP), Optimization, Routing problem, Scheduling, Collision Avoidance, Avoidance Constraints, Receding-horizon Formulation, Baseline Algorithm, Scalability, Mobile, Tablet, Web, Creative, Detail-oriented, Analytical Skills, Synthesize analysis, People management skills, Relationships, Effective campaigns, Technical/operational capabilities, Product development, Operations, Technology team, Project Management, End-To-End management, Campaign Workflow, Creative delivery, Campaign Performance, Key Metrics, KPI, CTR, Fill Rate, eCPM, CPA, CPG, Impression, Click, Lead scoring, Reach, ROI, Revenue, A/B Testing, Multivariate Testing, Path Analysis, HTML5, CSS, Flash, Web Responsive, SQL, Advanced Excel

---

# 1. Introduction

*This paper is a preprint of an extensive research soon to be described in a full paper.*
On Web Advertising over the last years, the Adv traffic kept growing. Due to this ceaseless increase of the number of Adv in the Web, Traffic Offices are becoming an important bottleneck of Adv Traffic. Hence, using decision support systems and optimization tools is more and more critical.
The Adv uploads play an important role in the Adv emissions and the workflow. The Adv creatives and tracking are also key components of the Adv Traffic Office. It also represents a non negligible part of trafficking cost. A better routing optimization allows to save costs on the Adv campaign management.

The *Trafficking Routing Problem* (*TRP*) consists in scheduling the management of Web Advertising (Adv) campaign between Trafficking campaigns without conflicts and in the most effective way. An incoming Adv has to be routed from its input acceptance to its staging area. While a delivering Adv has to be routed from its current incoming position to its delivering runway.
The Adv movements occur on a network of path of management that we call "adv-ways" which link Adv campaigns (see Figure 1) [1-12]. In practice this problem is issued by *Adv Traffic Controllers* (*ATCs*) on an operational window of typically 10 to 40 minutes.
The main constraints of the problem are related to the efficient Adv Loading: campaigns have to be separated from each other to avoid conflicts. Several other routing constraints must also be taken into account such as waiting times of the campaigns.
It is often difficult to design an optimum system of an Adv-way network. The Adv-way system may have a decisive influence on the capacity of the loading queue system, and thereby also the overall capacity of the Traffic Office. Considering that the load bearing strength of an Adv-way network represents an important item in the total investment costs. Therefore it is necessary to optimize the Adv-way network system layout to provide efficient wait for a campaign without undue expense.
The Adv-way network should permit safe, fluent and expeditious movement of the campaigns. They should provide the shortest and most expeditious connection of the loading queue with the staging area.
The safety of campaigns is enhanced if the Adv-way network is designed as one-way operation, and crossing other Adv-ways, and particularly loading queue, is minimized. In those Traffic Office where the number of campaigns movements during the peak hour traffic is relatively small, it is usually sufficient to provide only a short Adv-way to the loading queue to connect it to the loading of the campaign.

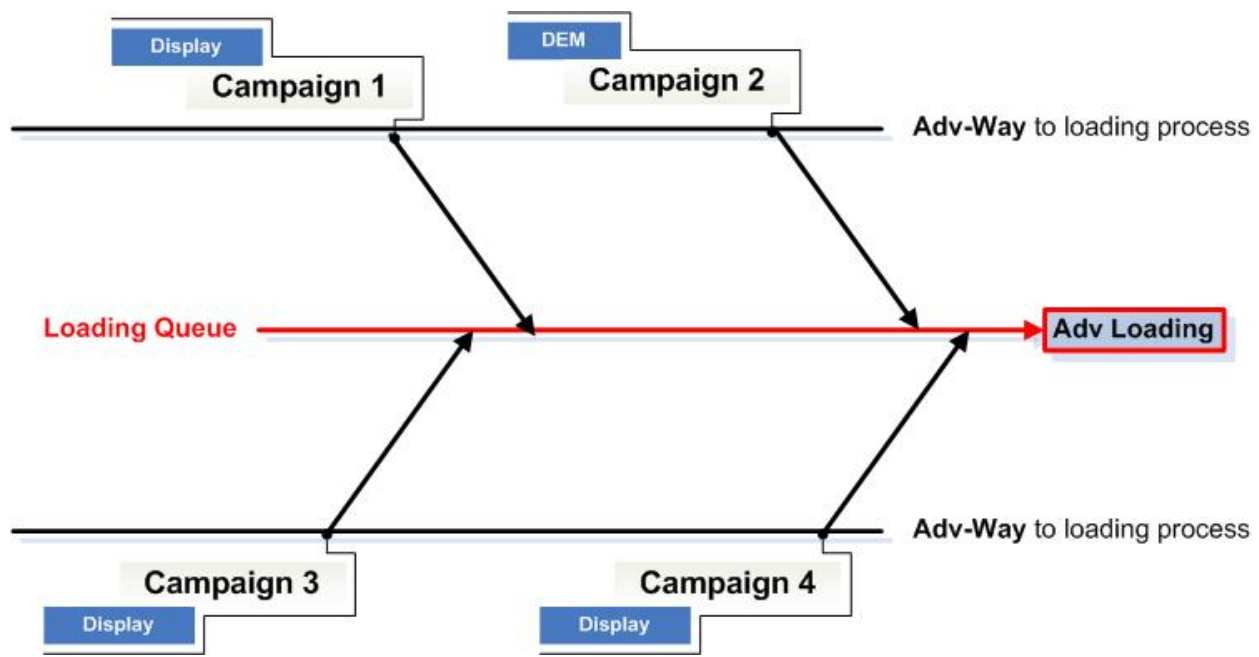

*Figure 1: Flow chart of a Traffic Office*

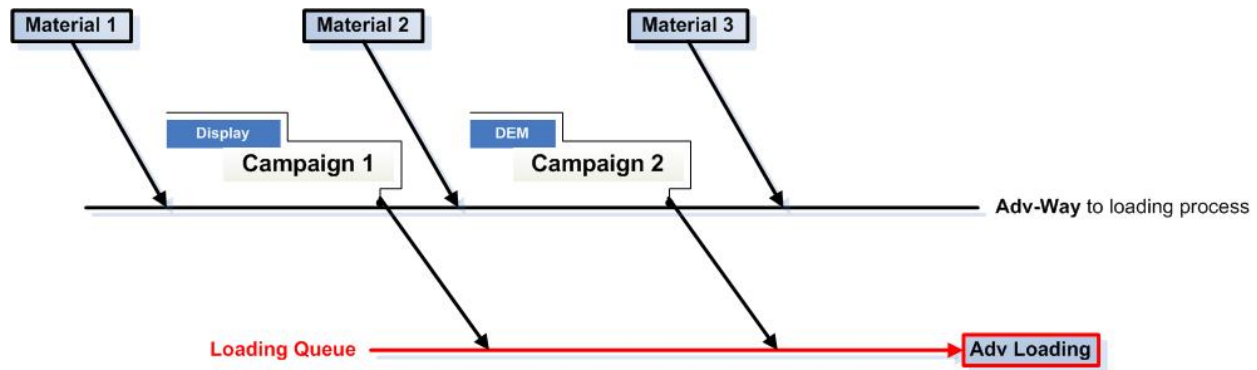

*Figure 2: Loading Queue*

Determination of the delivering runway length to be provided for the campaigns' workflow [1] is one of the most important decision in designing a Traffic Office. The runway length determines the types of campaign that may be used by the Traffic Office. The basic requirements for the runway parameters may be specified from market research into the types of campaigns to be delivered, the networks of the campaigns and prognoses of further market development at the Traffic Office in question.

The first step in specifying the necessary delivering runway length is to create a list of campaigns that may wish to use the Traffic Office. It is advisable to divide the campaigns into groups which are characterized by:
- type of materials required for the adv
- payload (monetary value for the adv)

Each of the campaigns groups requiring approximately the same runway length. In the majority of cases, only a small group of campaigns requires the longest class of delivering runway length, or even only one campaign: the "critical" campaign. The number of movements of critical campaign is sometimes so small that there is no economic justification for extending adv-way fulfils its requirements. In that case the operation of a critical type is usually still possible from the existing runway, though with an extra campaign cost, or reduced features of the campaign, or both at some times of the day and year. However, the Traffic Office will not use such a campaign if it would frequently require considerable limitation of the features of the campaign respect the cost.

The quality of a routing schedule can be defined by several Key Performance Indicators (KPIs). In this paper, we focus on four KPIs:
- the average *waiting time*,
- the average *completion time*,
- the average *delay* and
- the *On Time Performance* (*OTP*).

The *waiting time* measures the time an Adv spends on Traffic Office, between push back (i.e. leaving the incoming position) and take-off for loading. It includes any waiting time (e.g. queuing time due to others Adv management) and not just the time spent managing a single Adv.

The time consumption is not accurately known for the waiting process nowadays, but it mainly depends on the waiting. Other influencing factors have been identified, such as the number of stops, but their effects are less clear and of minor importance in comparison to the waiting time.

We are also interested in minimizing the *completion times*, i.e. the Adv upload times for departing Advs. In peak hours, the "adv-way" is often the main bottleneck of the Traffic Office. Minimizing upload times reduces the risk of "adv-way" starving and ensures a good use of its capacity. Minimizing waiting times reduces the Adv creativities and tracking time, which increases the quality of service.

In the Adv traffic industry, the main indicators of punctuality are the *average delay* per Adv and the *OTP*. The *average delay* is measured with respect to the scheduled times of waiting time and push back. For example, an arriving Adv creativity is 5 minutes late if the Adv creativity arrives at the staging area 5 minutes after the scheduled time and an uploading Adv is 5 minutes late if the Adv is pushed back from its staging area 5 minutes after the scheduled time. The *OTP L* is the percentage of Advs having a delay less than *L minutes*. The most common value of *L* in the industry is *15 minutes* and *OTP 15* is simply called *OTP*.

A common practice in Traffic Office is to push back Adv as soon as possible and to put it on hold to the "adv-way". It reduces the risk of "adv-way" starting and is beneficial for the upload delay and *OTP*. However, especially during peak hours, the "adv-way" capacity is often exceeded and a push back as soon as possible policy results in a take-off queue (see Figure 2). We call this process: *staging area holding*. Nevertheless, if an incoming Adv is held too long, it may not reach the "adv-way" in time for uploading and some "adv-way" capacity can be wasted. It may also prevent an arriving Adv from using the staging area (staging area blockage).

Accurate estimations of Adv ready times and waiting times are required to schedule the Adv loading adequately, i.e. holding staging area as much as possible in order to reduce waiting times, but without wasting the "adv-way" capacity. Accurate estimations of ready times are not always available in Traffic Office: those are often informed of an Adv ready time only when the Agent mails the Adv Order or the Adv creativity/tracking for push back and start up approval for Adv Loading. That is why the Traffic Offices designed the *Collaborative Decision Making* (*CDM*) project which main goal is precisely to improve predictability and information sharing between all stakeholders [10-13]. In a *CDM* Traffic Offices, Agents and Clients handlers are required to communicate and update an accurate "*ready time*" (typically 3 to 4 days in advance respect the online of the Adv). The ready time is called the *Target Off-Block Time* (*TOBT*) and corresponds to an estimation of the time at which the Adv creativity/tracking will be ready to push back on Adv Loading (all creativities are correct, tracking is correct, etc). In the Adv traffic industry, the push back scheduling is called often the "*pre-upload sequencing*".

In literature, it has been shown that staging area can reduce waiting times significantly without impacting the adv-way capacity (see *Section 2*). Nevertheless, the impact of staging area on the KPIs of the industry (*OTP* and delay) has not been investigated. In this paper, we propose a model including the *OTP* and delay indicators. We then address the following questions through a numerical study based on realistic instances from an Adv Traffic Office. How do the performance indicators compete ? Are the key indicators of the industry consistent with a sustainable development ? Can we propose better indicators ? Can we reduce waiting times by considering alternative paths ? What is the bottleneck in Adv-traffic routing operations ? The remainder of this paper is organized as follows.

A review of the literature and a summary of our contributions are presented in Section 2. In Section 3, we propose a model for the *TRP* and formulate it as a *Mixed Integer Linear Programming* (*MILP*) [5] in which only some of the variables are constrained to be integers, while other variables are allowed to be non-integers. We provide details on the data set and instances from a Traffic Office in Section 4. Then the results of our numerical study are given in Section 5. Finally, a conclusion and discussion of our results are presented in Section 6.

## 2. Paper contributions

The input of our model is the upload sequence of an Adv while the input of most of the models is the upload schedule with targeted upload times. We did this choice because manipulate sequences is more convenient than schedules for *ATCs*.

In that way, our model can be used as a tool for supporting adv-way sequencing decisions: it provides optimal upload times from a sequence, while accurately taking into account routing considerations.[13-14] Moreover besides the waiting time and completion time indicators we include the punctuality indicators of the industry (*OTP* and *delay*) in the objective function and we add staging blockage constraints.

In the *TRP* we consider the *OTP* and *delay*. We analyze the impact of including these KPIs in the optimization for a numerical study. We demonstrate that they are in contradiction with the objective to reduce waiting times in uploading Adv. As result we propose new indicators that are more sustainable, but also more relevant for stakeholders.

Our experiments show that the alternative paths method does not succeed in improving the KPIs significantly. That confirm the adv-way as the main bottleneck in Traffic Office.

## 3. Traffic Routing Problem formulation

In this section, we formulate the *TRP* as a *MILP* and we introduce the main notations. We present the model with a single path for each upload of an Adv.
For the proposed model, the main inputs are:
- the adv-way allocation,
- the adv loading,
- the scheduled incoming new campaign and
- the staging allocation plan (including the sequence of Adv upload operation for every staging area).

## 3.0 Mixed Integer Linear Programming MILP

*Mixed Integer Linear Programming* (*MILP*) formulations are widely used by exact solution methods in Operational Research. In comparison to *Linear Programming* (*LP*) formulations where the objective function and constraints all have to be linear, *MILP* formulations introduce an additional restriction of integrality for some variables. Unfortunately, since this restriction changes the nature of the search space from continuous to discrete, it often leads to problems which are much harder to solve, so that solution times for large problems may no longer be practical [15-22].

## 3.1. Single path model

### *Traffic network*
The *Traffic Network* is composed by three networks (*Material Network*, *Waiting Network*, *Ad-way Network*) and it's modeled as a graph $G = (N;E)$ with $N$ the set nodes and $E$ the set of edges [18-19]. There is a node for each intersection among the three networks and additional nodes in the Material Networks to represent the incoming materials. An edge represents an elementary segment of the network.
Overall the set of incoming and delivering campaigns is $C = C_{incoming} \cup C_{loading}$
For a campaign $i$, the single path from its origin *oi* to its destination *di* is $P_i = \left( o_i, u_2, ..., u_{|C_i|-1}, d_i \right)$

Let $N_i \subset N$, $E_i \subset E$ the set of nodes and edges that campaign $i$ can use. Note that the origin *oi* and the destination *di* are fixed by the adv-way and loading queue allocation (see *Figure 2*).

### *Adv characteristics*
An Adv campaign $i$ is ready to leave its origin *oi* at time *Toi*  For an Adv campaign $i \in C_{incoming}$, it corresponds to the *Target Off-Block Time* (*TOBT*) estimated by the Traffic Officer and the Agent.
The scheduled time for campaign $i$ is *SBi*, this time is used to measure the delay and the *OTP*.
For an incoming new campaign, it is the *Scheduled In-Block Time* (*SIBT*), i.e. the time a campaign is supposed to arrive at its adv-loaded allocation.
A campaign $i$ can spend a minimum (maximum) time *Tmin iuv* (*Tmax iuv*) on edge $uv \in E_i$
These times can be directly computed from the minimum and maximum times allowed to cross an edge *uv* for a campaign $i$ and from the edge length (see *Section 4*).
The adv-loading sequence is an input of our model. The position of departure $i \in C_{loading}$ in the take-off sequence is $\Gamma(i)$

### *Interactions between campaigns*
Campaigns *i* and *j* must have a minimum separation time at each node $u \in N_i \cap N_j$: if campaign *i* arrives first at node *u* at time *t*, then *j* cannot cross node *u* before *t+Siju.*

Let $G \subset C_{incoming} \times C_{loading}$ the set of possible staging blockages. A pair of campaigns *(i, j)* belongs to *G* if loading *i* and incoming *j* are assigned to the same staging and *i* is scheduled before *j* (in the staging allocation plan). In this case, loading *i* must leave the staging area before arrival *j* arrives in staging blockages.

### Decision variables
The main decisions in the single path approach is the time taken by the campaign to reach each node of the path. Our algorithm uses the following variables:

$t_{iu}$ : the time when campaign *i* reaches node $u \in N_i$. The origin time $t_{io_i}$ corresponds to the working time to load an Adv campaign. The loading time $t_{id_i}$ corresponds to the loading time for an incoming new campaign.

$\beta_i = 1$ if campaign *i* is delayed by more than $L \geq 0$ with respect to the scheduled reference time *SBi* (if $\delta_i > L$), 0 otherwise.

$\delta_{iu}$ : the delay of campaign *i* to its scheduled reference time *SBi*. The delay is $\max(0, t_{id_i} - SB_i)$ for an incoming campaign while for a loaded campaign is $\max(0, t_{io_i} - SB_i)$

$z_{iju} = 1$ if campaign *i* arrives before campaign *j* in node $u \in N_i \cap N_j$, 0 otherwise.

### Objective function
We aim in minimizing the following performance indicators.

$\sum_{i \in C} \beta_i$         : number of campaigns delayed by more than L

$\sum_{i \in C} \delta_i$         : total time delay

$\sum_{i \in C} (t_{id_i} - t_{io_i})$ : total waiting time

$\sum_{i \in C} (t_{id_i} - T_{O_i})$ : total completion time

### MILP formulation for Single path model
The single path problem can be formulated by *Mixed Integer Linear Programming* (*MILP*) define in (1) and (2) and the relative constraints (3)

The following objective function is a linear combination of the above four indicators using non negative coefficients : $k_{OTP}, k_{delay}, k_{wait}, k_{ct}$

Indexing the coefficients we can differentiate campaigns: as a campaign waiting to be uploaded will block the uploading of other campaigns.

$$(1) \quad \sum_{i \in C} k_{OTP,i} \, \beta_i + \sum_{i \in C} k_{delay} \, \delta_i + \sum_{i \in C} k_{wait} \, (t_{id_i} - t_{io_i}) + \sum_{i \in C} k_{ct} \, (t_{id_i} - T_{O_i})$$

Constraints (2) ensure the definition of delay variables $\delta_i$ and OTP variables $\beta_i$.
Constraints (3) specify the domains of the variables.

$$(2) \quad \delta_i \geq t_{iO_i} - SB_i \qquad \forall i \in C_{loading}$$

$$\delta_i \geq t_{iO_i} - SB_i \qquad \forall i \in C_{incoming}$$

$$\delta_i \leq L + M \beta_i \qquad \forall i \in C_{loading}$$

$$(3) \quad t_{iu} \geq 0 \quad \forall i \in C, \forall u \in N_i$$

$$\delta_i \geq 0 \quad \forall i \in C$$

$$\beta_i \in \{0,1\} \quad \forall i \in C$$

$$z_{iju} \in \{0,1\} \quad \forall i \neq j \in C, \forall u \in V_i \cap V_j$$

Note that generalizing the coefficient it is possible to minimize the number of campaigns delayed by more than *L* that is equivalent to maximizing the percentage of campaigns with a delay less than *L*, that we define as *OTP L*

$$(4) \quad \min k_{OTP} \sum_{i \in C} \beta_i + k_{delay} \sum_{i \in C} \delta_i + k_{wait} \sum_{i \in C} \left(t_{id_i} - t_{iO_i}\right) + k_{ct} \sum_{i \in C} \left(t_{id_i} - T_{O_i}\right)$$

### *Bounding constraints*
Constraints (5) ensure that new campaigns cannot push back before they are ready to.

Constraints (6) ensure that incoming new campaigns start waiting as soon as they arrive to the Traffic Office, in order to free the adv-way.

$$(5) \quad t_{iO_i} \leq t_{jl_j} \quad \forall (i,j) \in G$$

$$(6) \quad t_{iO_i} = T_{O_i} \quad \forall i \in C_{incoming}$$

### *Staging blockage constraints*
Constraints (7) ensure that an incoming campaign does not be loaded until the previous campaign has ended the loading (staging blockage constraints).

$$(7) \quad t_{iO_i} \geq T_{O_i} \quad \forall i \in C_{loading}$$

### *Time constraints*
Constraints (8) ensure the respect of minimum and maximum time spent on every edge (time constraints). The maximum time spent on an edge allows to prevent adv campaign from stopping in certain adv-way segments (e.g. adv-way crossing). It also ensures that the capacity of the edge is not exceeded (i.e. no more campaign that its length allows it).

$$(8) \quad T_{\min,iuv} \leq t_{jv} - t_{ju} \leq T_{\max,iuv} \quad \forall i \in C, \forall uv \in E_i$$

### *Runway sequencing constraints*
Constraints (9) ensure the definition of sequencing variables *ziju*, i.e. either campaign *i* arrives before campaign *j* in node u 2 Vi \ Vj or the opposite.
Constraints (10) ensure that the campaign upload sequence is respected.

$$(9) \quad z_{iju} + z_{jiu} = 1 \quad \forall i, j \in C, \forall u \in N_i \cap N_j$$

$$(10) \quad z_{iju} = 1 \quad \forall i, j \in C_{loading}, d_i = d_j = u$$

### Separation constraints

Constraints (11) prevent campaign from bumping into each other at every node with a priority conflict (see *Figure 3*), where *K* is supposed to be a high enough value (e.g. 10 times the time window is largely sufficient, it remains in forcing every campaign to end waiting to be uploaded in less than 10 times the time window which is reasonable).

$$(11) \quad t_{ju} \geq t_{iu} + S_{iju} - K(1 - z_{iju}) \quad \forall i, j \in C, \forall u \in N_i \cap N_j$$

### Overtake and Head-on constraints

Constraints (12) prevent two campaigns to be done concurrently - from using an edge in opposite directions simultaneously (see *Figure 3*). Constraints (9) also prevent a campaign from overtaking another one on an edge, which is impossible (see *Figure 3*).

$$(12) \quad z_{iju} = z_{ijv} \quad \forall i, j \in C, \forall uv \in E_i \cap E_j$$

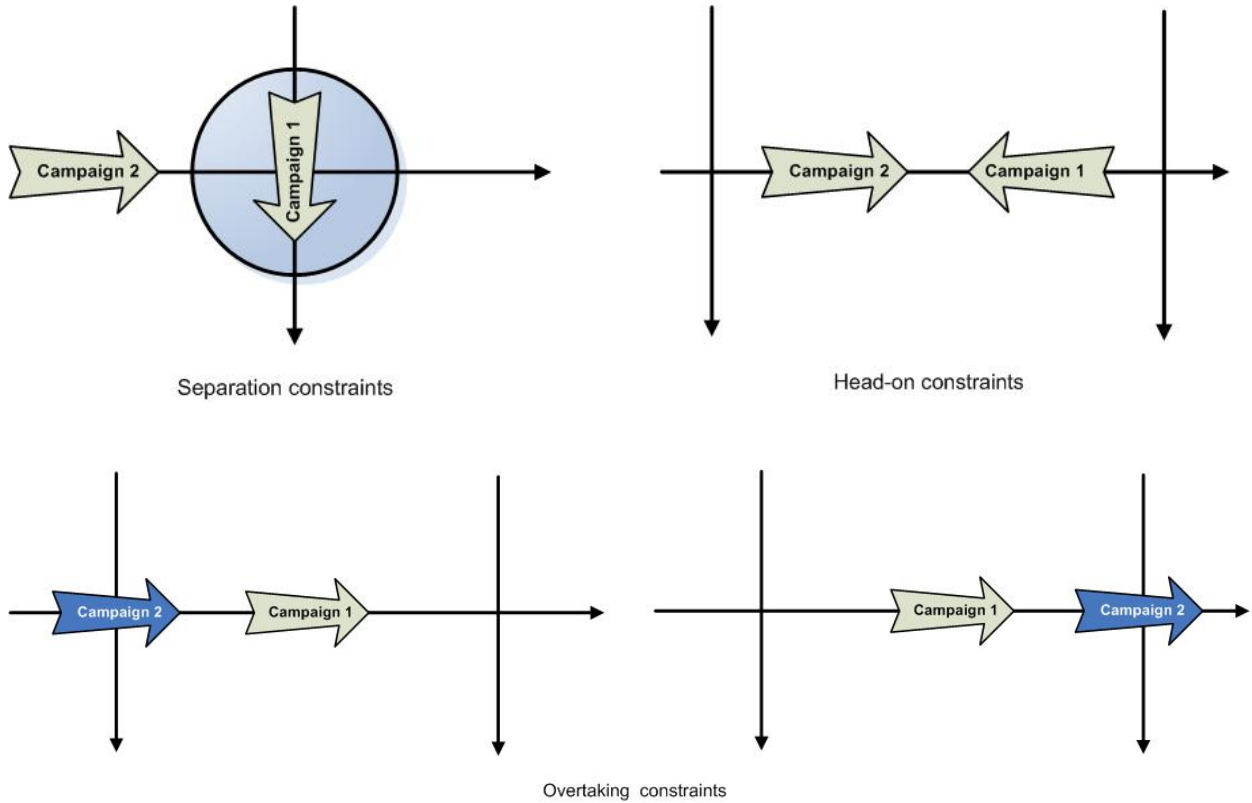

*Figure 3: Safety constraints*

## 4. Instances for an operational day

In this section we present our instances and how they were generated. Each instance represents an operational day in a Traffic Office.
For our own interest we did that independent research projects also through specific innovative tasks validate towards average working time declared on "specification required" by the main worldwide industry leading Advertising Agency.
*http://www.sizmek-sea.com/Spec*
*http://support.adform.com/...specifications/general-specifications*
*http://creative-weborama.com/category/uncategorized*
*http:// groupm.dk/Banner-sizes-and-specifications*

As discussed in the *Introduction* the campaigns were divided into two groups characterized by: type of materials required for the adv and payload (monetary value for the adv). Relative data about those metrics was desumed on "specification required" by the main worldwide industry leading Advertising Agency mentioned above.

### *Delivering runway configuration.*
The most frequent delivering campaign in a Traffic Office is the Display with start on a working day night and stop after one day.
In the month of September 2016, we have selected 8 busy days (with more than 800 campaigns) in which more than 70% of campaigns were operated in this adv-way configuration. The average number of incoming Adv new campaign and loaded campaign by hour is presented in *Figure 4*. Minimum delivering runway separation times used at the Traffic Office are presented in *Table 1*.

An edge represents an elementary segment of the three linked networks. A node needs to be defined for each intersection. There is also a node for each campaign-delivering.
The graph is composed of:
- 28 nodes (Materials Network: 11 nodes, Waiting Network: 10 nodes, Adv-Way Network: 7 nodes),
- 35 edges (Materials Network: 12 edges, Waiting Network: 12 edges, Adv-Way Network: 11 edges).

The standard path between each campaign-delivering and each adv-way was provided by the Traffic Office, as well as the standard push back scheme and its duration, for each campaign-delivering. We observed that standard paths were used for more than 83% of campaigns.
Campaign flow data around the Network also provides an estimation of the maximal time spent. Based on these data, we assume a maximal time of 15 days for the campaigns around the Adv-way Network (in blue in *Figure 5*), of 5 days for the campaigns on Material Network (in red in *Figure 5*) and of 10 days for the other Waiting Network.

To ensure the respect of minimum and maximum time spent on every edges we assume a minimum time of 2 days. The minimum separation time ($Si_{ju}$) between two campaign is assumed to be 40 seconds for every nodes (except the runways, see *Table 1*).

### *Campaigns*
Direct information on campaigns are not delivered with the campaign-flow but only records composed of campaign identifier, position in the Traffic Office and time stamp while the Traffic operational database provides other useful data for each campaign.

It gives the *Scheduled In-Block Time* (*SIBT*), denoted by $SB_i$ in our model. For each arriving campaign, it also provides the *Actual Incoming Time* (*AIT*) which can be used to define the release time $T_{oi}$.
As a push back as soon as possible policy is used in the Traffic Offices most of the time, we have decided to take the *Actual Off-Block Time* (*AOBT*) to define the release date $T_{oi}$ for departing campaigns. Finally, the take-off sequence is the actual one which can be derived from the *Actual Take-Off Times* (*ATOTs*) [22-26].

*Average performance indicators.*
All results are averaged among the campaign for 5 working days. For instance, an average waiting time of 10 minutes means that it takes on average 10 minutes for a campaign to stay in the waiting area, among the 5 working days. We choose the *OTP 10* indicator ($L = 10$) as punctuality indicators.
*Networks for Adv-Way, Waiting and Materials. Figure 5* presents the graph of the Adv-way Network with the Waiting Network and the Materials Network.

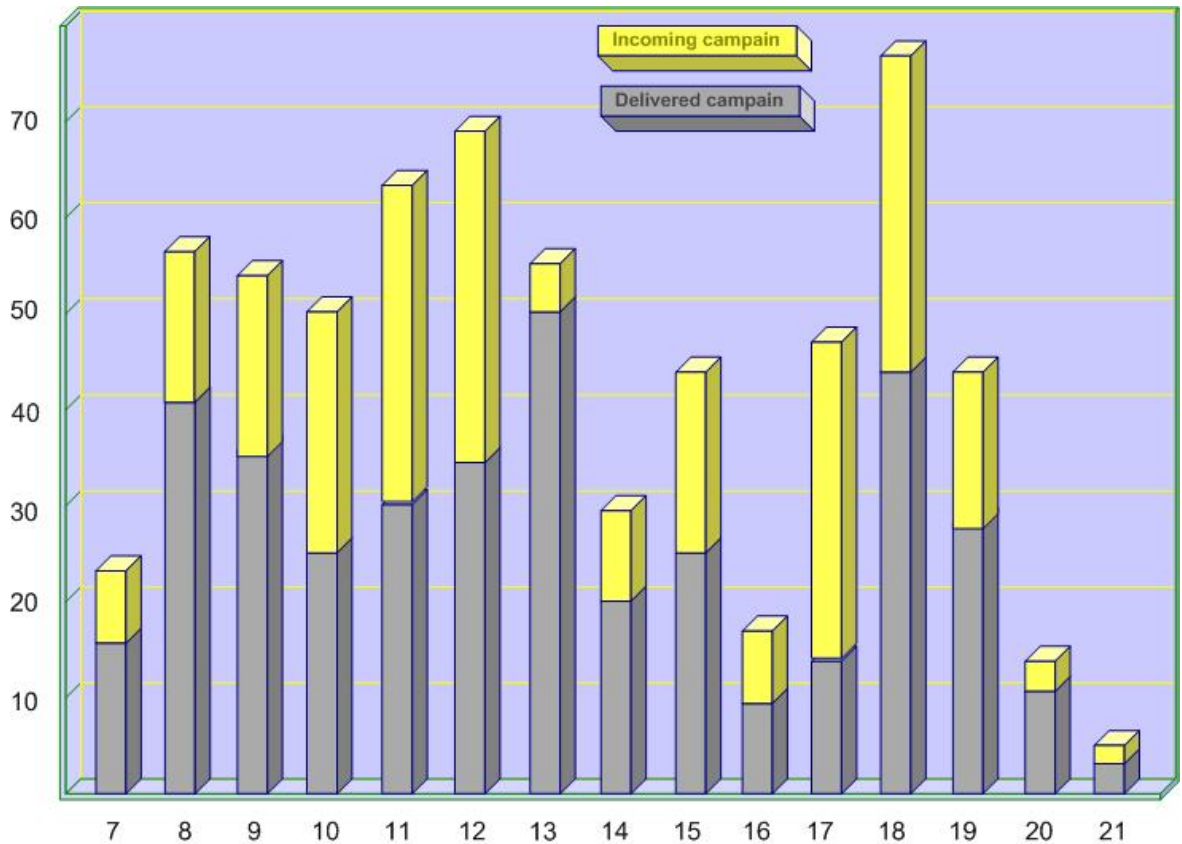

Figure 4: Average profile of instances

|  |  | Loading campaign | | |
|---|---|---|---|---|
| Time [s] |  | S | M | L |
|  | S | 90 | 120 | 120 |
| Incoming | M | 60 | 60 | 90 |
| campaign | L | 60 | 60 | 60 |

Table 1: Minimum adv-way separation time at Traffic Office (S = short time, M = medium time and L = long time)

## 5. Results on quantitative data

The results are presented for the single path problem, except in *Section 5.3* where we study the effect of the number of paths [16]. We set *k.wait = 1* without loss of generality, as we vary the other weights *k.ct*, *k.delay* and *k.OTP*. Results of mixed integer programs were obtained with C.plex 12.4 solver using default parameter tuning on a personal computer (Intel Core i7 Extreme 3.70Ghz, 16Go RAM) under Ubuntu 14.04 LTS operating system. Java Concert API was used to define the models.

### 5.1. Sliding window optimization

It is not possible to schedule the movements of a campaign for the entire day as there are many stochastic events. Usually in literature and practice a routing problem – as the *Trafficking Routing Problem* (*TRP*) – is solved dynamically with a sliding window approach. With large time window a better solutions on *TRP* can be found but the computation times increase too.

The optimization does not need to be performed continuously but only when a new campaign enters the system. Once a campaign has started to be worked by the Traffic Office, changing its schedule is not allowed in the next time windows, but it has to be taken into account to ensure a conflict free routing.
In the rest of the numerical study we set a time window of 30 minutes. This assumption seems reasonable in the context of a usual approach in which Clients and Traffic Office are required to communicate accurate ready times 3 to 4 days in advance.
With a 30 minutes time window, computation times were always below two seconds for the single path approach. It appears that a time window of 15 minutes is sufficient in our test case, i.e. longer time windows do not provide better solutions. This value may be Traffic Office dependent and cannot be generalized without experiments in other Traffic Offices.

### 5.2. Including the punctuality key performance indicators

We consider the *OTP* and the *average delay* as main punctuality indicators for campaigns and Traffic Offices. However the client generally focuses on waiting time and completion time indicators. In this section, we study the impact of including the average delay and the *OTP* in the optimization.

#### 5.2.1. Average delay

*Figure 6* presents the effect of the weight *k.delay* on all KPIs for incoming campaigns (dashed lines) and delivering campaigns (solid lines) with different values of *k.ct*.
For arrivals, we observe that KPIs are not much impacted by *k.delay*, which can be explained as follows. The contribution of a delayed arriving campaign *i* to objective function (1) is fixed (see constraints (3)) because the loading time is fixed.

$$i \in C_{arr} \left( t_{id_i} > SB_i \right)$$

$$(13) \quad \left( k_{delay} + k_{wait} + k_{ct} \right) t_{id_i} + k_{OTP} 1_{\{t_{id_i} > SB_i + L\}}$$

We consider the variables *tidi* for the KPIs because including the delay adds redundancy been correlated with the waiting time.

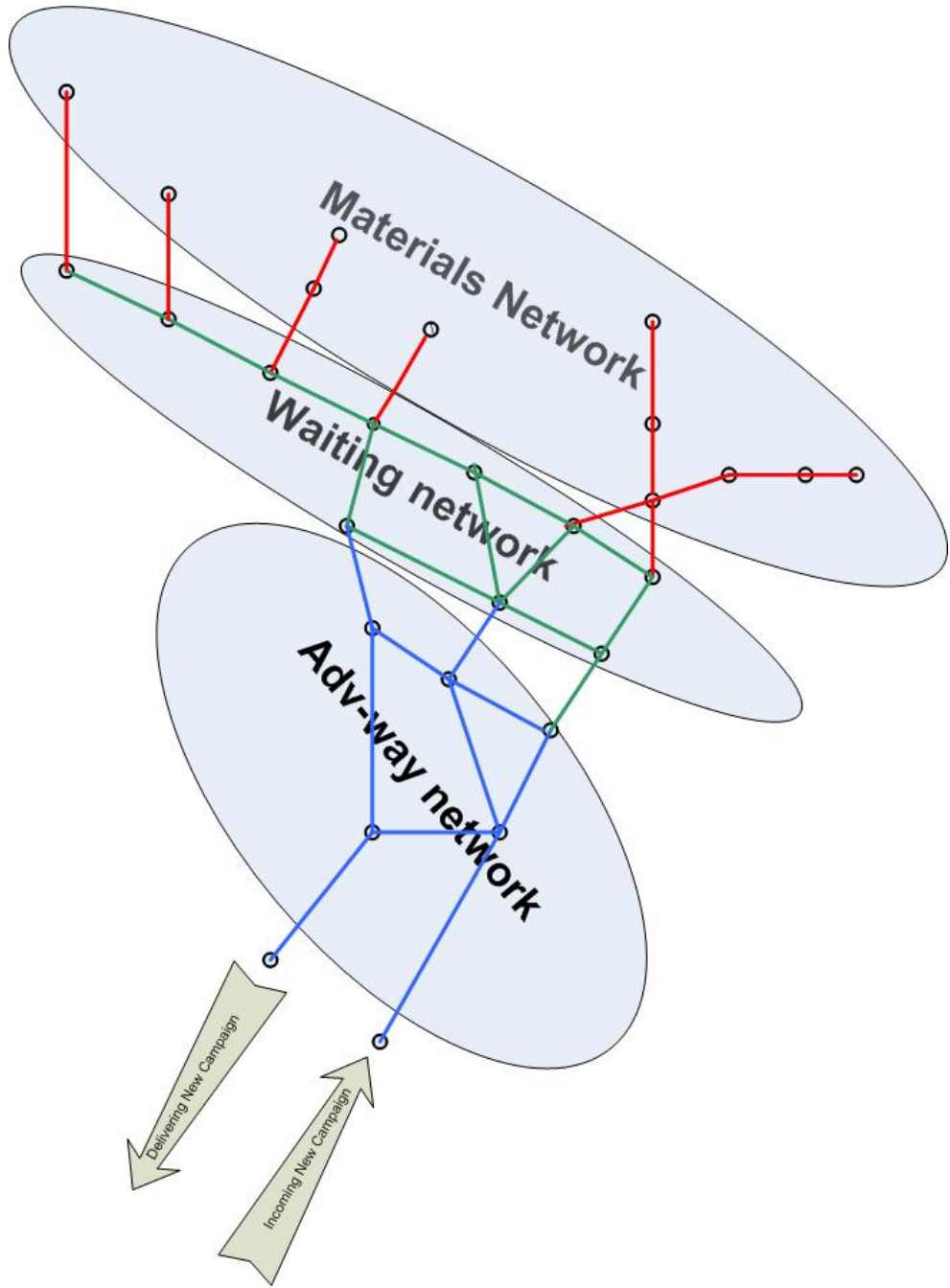

*Figure 5: The Adv-way network, Waiting Network and Materials Network*

While for new campaign, increasing *k.delay* reduces delays but increases waiting times.
When *k.delay = k.wait* we can observe a threshold effect between the delay and the waiting time due to the contribution of a delayed departure *i 2 Fdep (tioi > SBi)* to the objective function, which is, within a constant,

$$(14) \qquad \left(k_{delay} - k_{wait}\right)t_{iO_i} + \left(k_{wait} - k_{ct}\right)t_{id_i} + k_{OTP} 1_{\{t_{iO_i} > SB_i + L\}}$$

It clearly highlights an opposition of the waiting time and the delay for delivering the campaign.

When *k.delay – k.wait > 0*, pushing back campaign earlier (which reduces *t.ioi*) is preferable as it reduces delays. But it leads to longer waiting times when the loading queue is congested.

When *k.delay – k.wait < 0*, holding campaign at staging areas as much as possible (which increases *t.ioi*) is more profitable and avoids loading queuing. It consequently decreases waiting times, but implies larger delays.

To further illustrate the opposition between the waiting time and the delay for loading the campaign, *Figure 7* details the results along the day with a 30 minutes time window and a 5 minutes step. For instance, at 9h05, *Figure 7(a)* plots the number of a loading campaign in the time interval [9h05,9h35]. *Figure 7* plots the additional waiting time and the additional delay when instead of *k.delay = 0* we set *k.delay = 2*

In lows hours campaign go to the loading queue in the shortest time to be put online immediately and all performance indicators are optimized.
While during the peaks hours, the loading queue is saturated and campaigns cannot be put online as soon as they reach the campaign queue. They must either wait at staging areas or at the loading queue. When *k.delay = 0*, staging holding is preferred since it reduces the waiting time. When *k.delay = 2*, pushing-back earlier is preferred in order to reduce delays to the scheduled push back time.

In conclusion, delays cannot be significantly reduced without degrading waiting times in peak hours.

### 5.2.2. New delivered punctuality indicators

Staging holding succeeds in reducing the waiting time significantly by transferring loading queuing time without materials for the campaign on to the staging areas. We observed in *Figure 7* that it is particularly efficient during delivery peaks of all the campaigns. Nevertheless, our analysis also shows that this practice degrades the punctuality indicators. Hence Traffic Office may be reluctant to use staging holding and may prefer a push back ASAP policy to ensure good departure indicators. In this section, we question the relevance of *OTP* and delay indicators for campaigns of a Traffic Offices and propose new punctuality indicators.
For Traffic Offices measuring the punctuality with respect to push back times is not accurate as additional delays occur during the waiting process and particularly in the loading queue.
Moreover campaigns are not accountable for the delay between the ready time and the push back time. Consequently an idea could be to base the measure of loading punctuality for campaigns on the ready time and not on the push back time.
We propose to define the *Schedule Loading Time* (*SLT*) as *Scheduled In-Block Time* (*SIBT*) plus a constant depending on the Traffic Office, for instance the average loading completion time. Our models can easily be adapted to measure the *delay* and the *OTP* with respect to *SLT*. Constraints (10) can be merged with constraints (11) as follows :

$$t_{id_i} \leq SB_i' + L + M \beta_i \qquad \forall i \in C$$
$$\delta_i \geq t_{id_i} - SB_i' \qquad \forall i \in C$$

where *SB$_{0i}$* is the *Scheduled In-block Time* (*SIBT*) for incoming new campaign and the *Schedule Loading Time* (*SLT*) for delivered new campaign. *OTP* constraints (12) are unchanged.

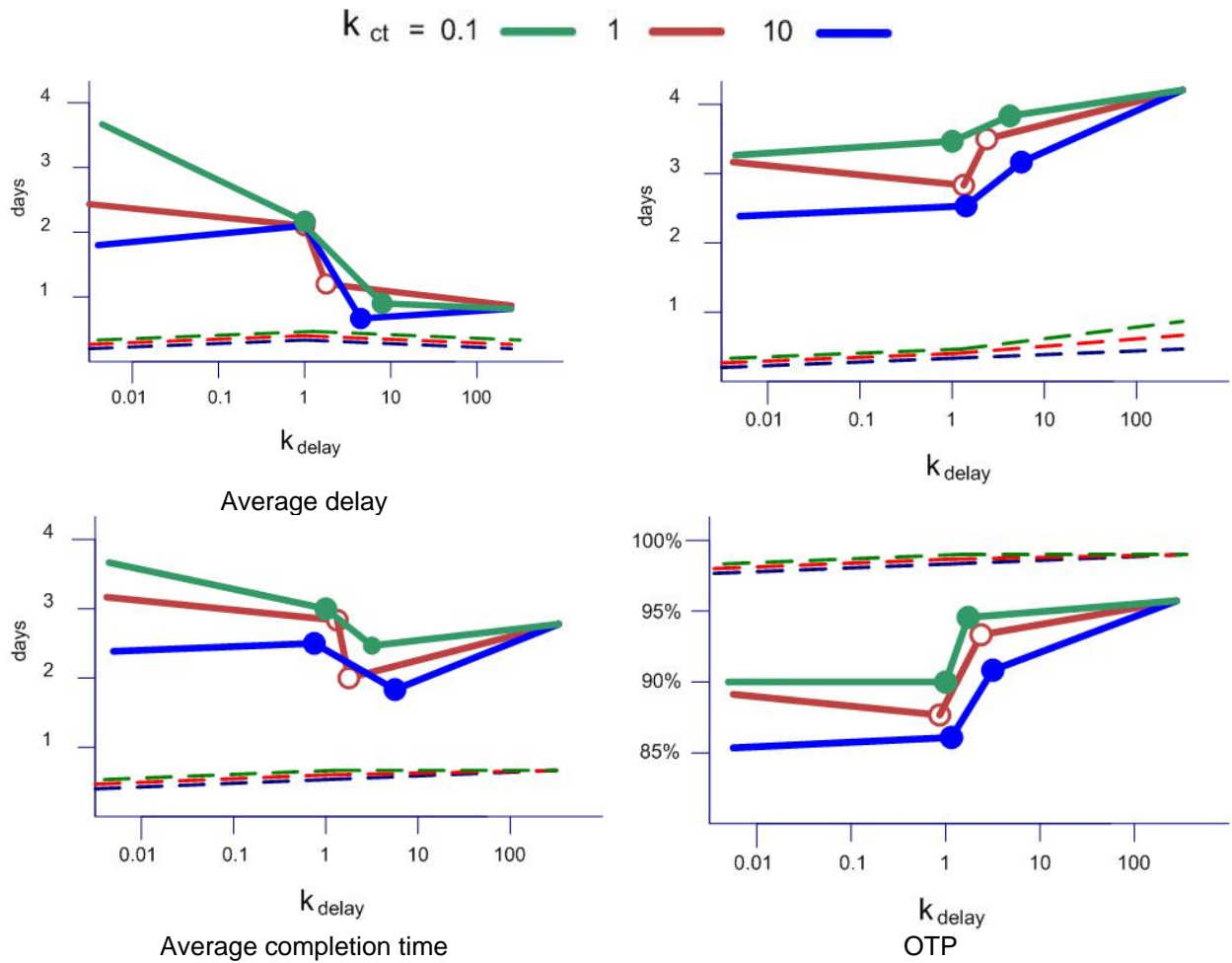

*Figure 6: Effect of including the delay indicator in the optimization for incoming new campaign (dashed lines) and departures (solid lines) (k.wait = 1, k.OTP = 0)*

## 5.3. Bottleneck (constraint) analysis

A bottleneck (or constraint) in a supply chain determine the throughput of a supply chain, it means the resource that requires the longest time in operations of the supply chain for certain demand.
The adv-way network can be divided in three distinct parts: the runway, the staging areas and the adv-way area. In this section, we evaluate the impact of each area on the routing, by relaxing its constraints in the optimization.

*Figure 8* presents the results of this analysis.
*All constraints* means that all constraints are taken into account.
*Adv-way* means that constraints of the adv-way area are relaxed.
*Stage* means that constraints of the stage area are relaxed.
*Runway* means that separation constraints of the runway are relaxed.

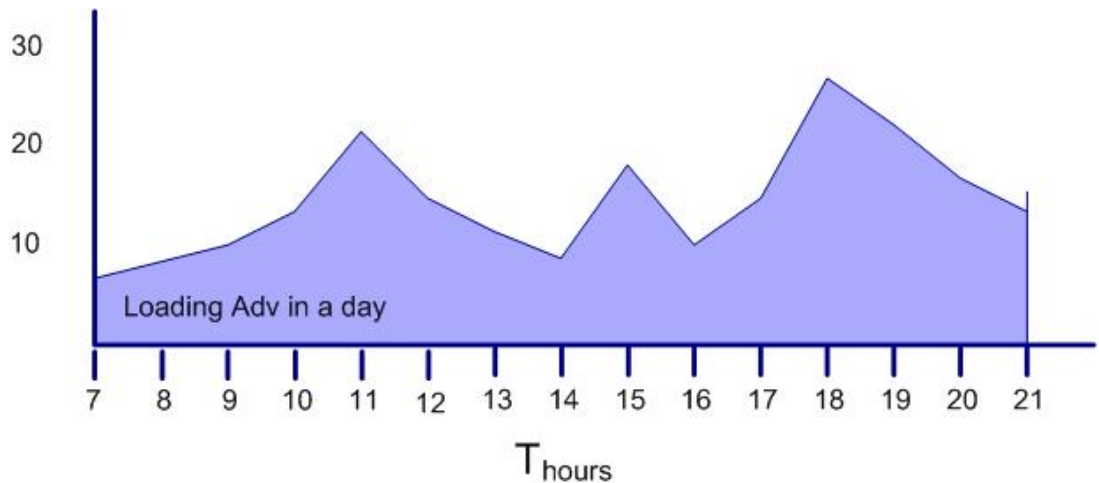

(a) Profile

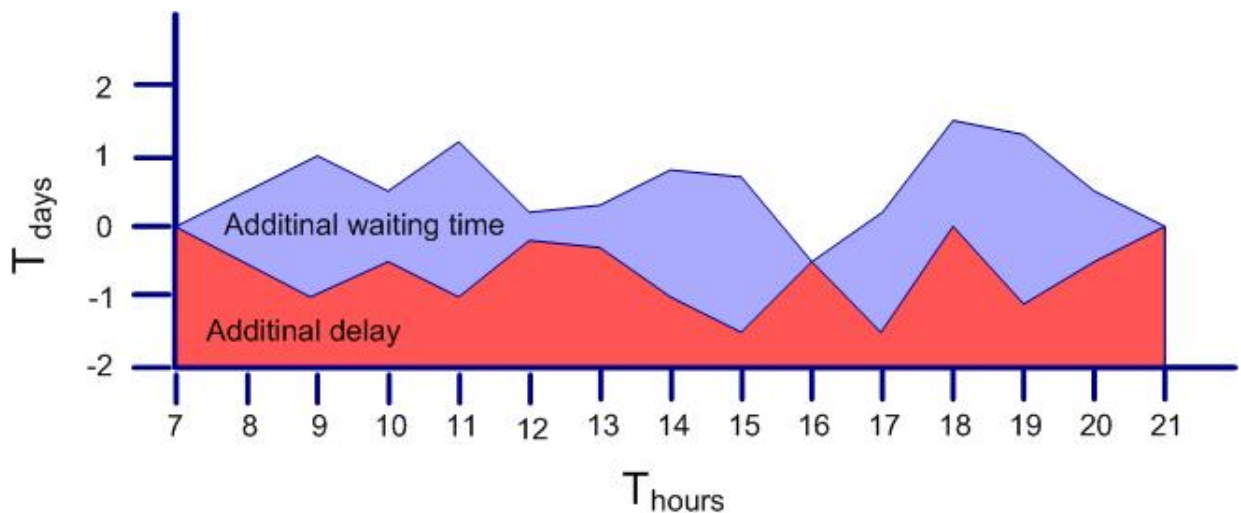

(b) Additional waiting time and delay when *k.delay = 2* instead of *k.delay = 0*

*Figure 7: Effect of including the delay indicator in the optimization along the day (departures only, k.ct = 2, k.wait = 1)*

Besides no loading adv sequence is forced. No constraint is the case where all the above constraints are relaxed and campaign stage at minimum time without stopping anywhere. The delay and the OTP are measured with respect to *SIBT / SLT* (as defined in *Section 5.2.2*).

*Figure 8* shows that the adv-way is the main bottleneck: all indicators are consequently impacted, particularly the completion time, the delay and the *OTP*. It also shows that the staging area has a limited impact on indicators, which join the conclusion of previous section.
We also observe that the stage area impacts all indicators in a significant way.

The lack of benefits provided by the alternate path can be explained by the structure of the Waiting Network in *Figure 5*.
Most of head-on conflicts between incoming and delivered new campaigns are avoided because there are two main parallel adv-ways serving the staging areas and each one can be used in a different direction in the single path approach.

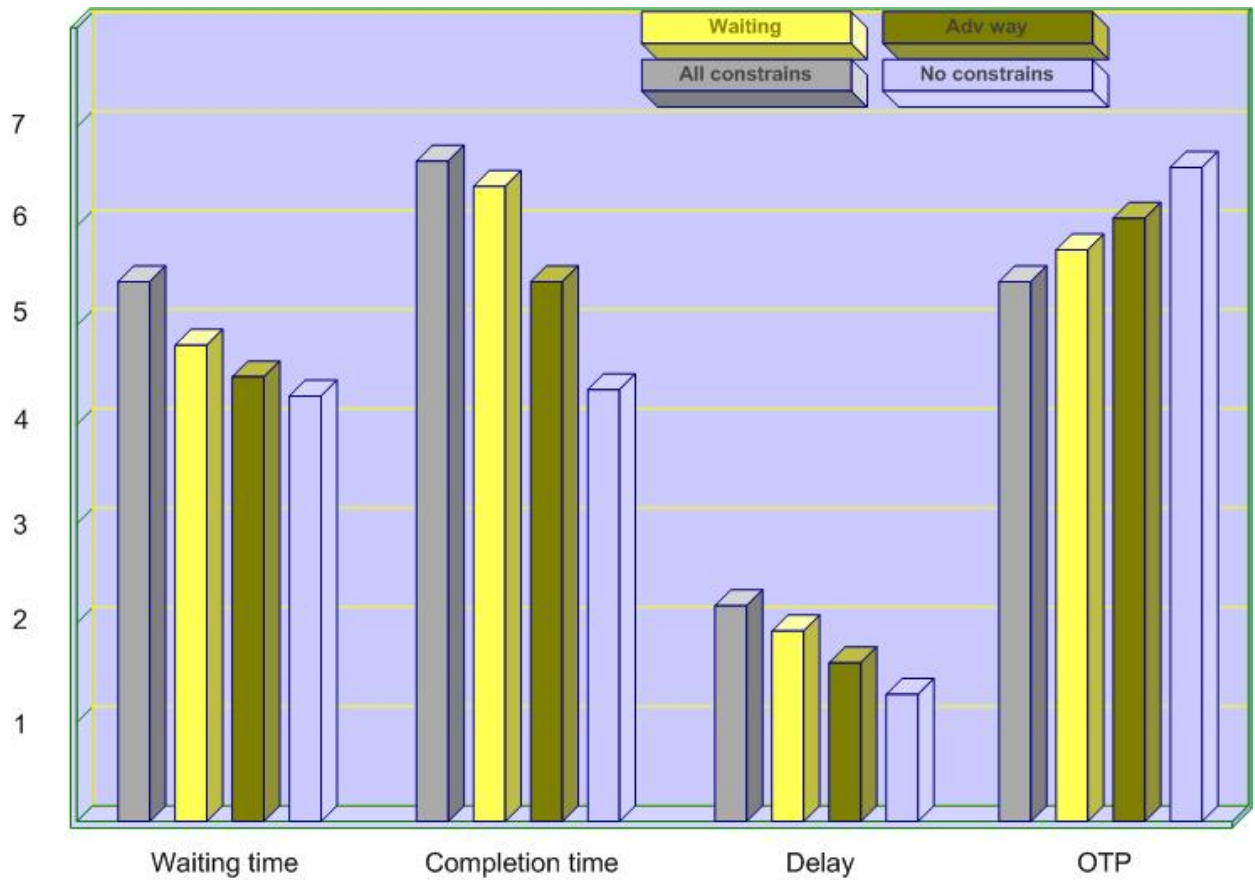

*Figure 8: Bottleneck analysis (k.ct = 2; k.wait = 1; k.OTP = k.delay = 0, new OTP and delay)*

While the area around the staging areas is much more intricate and generally offers a single adv-way. So the alternative paths model do not allow to avoid much more conflicts that the single path model. This intuition is further explored in the next section through an analysis of the Traffic Office bottlenecks.

## 6. Conclusion

In this paper, we formulate the *Trafficking Routing Problem* (*TRP*) as a *Mixed Integer Linear Programming* (*MILP*). We present a formulation with a single path. Our model takes into account the classical performance indicators (average delay and *OTP 10*).

In a numerical study based on data from a virtual Traffic Office (validate towards average working time declared on "specification required" by the main worldwide industry leading Advertising Agency), we first show how the punctuality indicators are in contradiction with a sustainable management of Traffic Office. The punctuality of departures is currently measured with respect to push back times, which encourages to push back as soon as possible and results in large waiting times in peak hours because of adv-way congestion.
Including the delay in the objective function leads to a waiting time increase of 1 minute in average for departures at the Traffic Office. In more congested situations, this increase can reach 6 minutes. Including the *OTP 10* in the objective function has less impact in current traffic situations. However, in more congested situations, it also leads to longer waiting times.

We propose to measure the punctuality of Adv Traffic Office with respect to adv loading times and not with respect to push back times. We show that this new measure of punctuality do not prevent staging area holding. Besides they are more appropriate since they capture additional delays between push back and adv loading.

We also show that the adv way is the main bottleneck of a Traffic Office and that considering alternative paths do not improve the performance indicators significantly.

Numerical experiments were performed in a Traffic Office and we may wonder to what extent our results can be generalized to Traffic Offices.

In congested Traffic Offices, the *delay* and *OTP* indicators will intuitively not be adequate to measure punctuality, as they encourage to ask for adv materials as soon as possible and lead to long waiting times. In non congested Traffic Office, it will not matter as an asking for adv materials as soon as possible policy should be nearly optimal.

The main parallel adv ways serving the "waiting areas" in a Traffic Office prevent most of head-on conflicts in the adv way area. Such a structure is very common and is used in the most frequented Traffic Offices in the world. In such configurations, the alternate path approach will probably not bring much with respect to the single path approach. However, the alternate path approach is certainly more beneficial in Traffic Offices with more complex adv way layout, typically with adv way crossing.

**Massimiliano Dal Mas** is an engineer working on webservices, trafficking and online advertising and is interested in knowledge engineering. In the last years he had to play a critical role at Digital Advertising business, cultivating relationships with key publisher partners with experience managing a team. Been responsible for all day to day operations with partners and consult on the best ways to monetize their properties. His interests include: user interfaces and visualization for information retrieval, automated Web interface evaluation and text analysis, empirical computational linguistics, text data mining, knowledge engineering and artificial intelligence. He received BA, MS degrees in Computer Science Engineering from the Politecnico di Milano, Italy. He won the thirteenth edition 2008 of the CEI Award for the "best degree thesis" with a dissertation on "Semantic technologies for industrial purposes" (Supervisor Prof. M. Colombetti). In 2012, he received the "best paper award" at the IEEE Computer Society Conference on Evolving and Adaptive Intelligent System (EAIS 2012) at Carlos III University of Madrid, Madrid, Spain. In 2013, he received the "best paper award" at the ACM Conference on Web Intelligence, Mining and Semantics (WIMS 2013) at Universidad Autónoma de Madrid, Madrid, Spain. His paper at 2013 W3C Workshop on Publishing using CSS3 & HTML5 has been appointed as "position paper", Paris, France.